\documentclass[runningheads]{llncs}
\usepackage{amsmath,amssymb,amsfonts}
\usepackage{algorithmic}
\usepackage{cite}
\usepackage{graphicx}
\usepackage{textcomp}
\usepackage{booktabs}
\usepackage{multirow}
\usepackage{xcolor}
\usepackage{color, colortbl}
\usepackage{xurl,lineno,microtype}
\usepackage[hidelinks]{hyperref}

\def\BibTeX{{\rm B\kern-.05em{\sc i\kern-.025em b}\kern-.08em
    T\kern-.1667em\lower.7ex\hbox{E}\kern-.125emX}}

\definecolor{Gray}{gray}{0.9}

\begin{document}

\title{Eclipse Qrisp QAOA: description and preliminary \\ comparison with Qiskit counterparts}

\titlerunning{Eclipse Qrisp QAOA}

\author{Eneko Osaba\inst{1} \and
Matic Petri\v{c}\inst{2} \and
Izaskun Oregi\inst{1,3} \and
Raphael Seidel\inst{2} \and \\
Alejandra Ruiz\inst{1} \and
Sebastian Bock\inst{2} \and
Michail-Alexandros Kourtis\inst{4}
}

\authorrunning{Osaba et al.}

\institute{TECNALIA, Basque Research and Technology Alliance (BRTA), Derio, Spain \and
Fraunhofer Institute for Open Communication Systems (FOKUS), Berlin, Germany \and
European University of Gasteiz, EUNEIZ, Vitoria-Gasteiz, Spain \and
National Centre for Scientific Research “Demokritos”, Agia Paraskevi, Greece\\
\email{eneko.osaba@tecnalia.com}}

\maketitle

\begin{abstract}
This paper focuses on the presentation and evaluation of the high-level quantum programming language \texttt{Eclipse} \texttt{Qrisp}. The presented framework, used for developing and compiling quantum algorithms, is measured in terms of efficiency for its implementation of the Quantum Approximation Optimization Algorithm (QAOA) Module. We measure this efficiency and compare it against two alternative QAOA algorithm implementations using IBM’s Qiskit toolkit. The evaluation process has been carried out over a benchmark composed of 15 instances of the well-known {\it Maximum Cut Problem}. Through this preliminary experimentation, \texttt{Eclipse} \texttt{Qrisp} demonstrated promising results, outperforming both versions of its counterparts in terms of results quality and circuit complexity.

\end{abstract}

\section{Introduction} \label{sec:intro}

The advent of quantum technologies is poised to play a pivotal role in different industries in the near future. Quantum Computing (QC), which leverages the principles of quantum mechanics to process information, is continuously making advances to bridge quantum processing with real-world applications. Despite being a not yet fully mature technology, this paradigm is generating great interest in the scientific community. 

Two types of quantum computers coexist: quantum annealers and gate-based devices. This work is focused on gate-based computers for solving optimization problems. In this context, Variational Quantum Algortihms (VQA) are the most widely applied optimization methods. As explained in \cite{cerezo2021variational}, ``\textit{the trademark of VQAs is that they use a quantum computer to estimate the cost function of a problem (or its gradient) while leveraging the power of classical optimizers to train the parameters of the quantum circuit}".

\begin{figure}[t]
	\centering
	\includegraphics[width=0.70\linewidth]{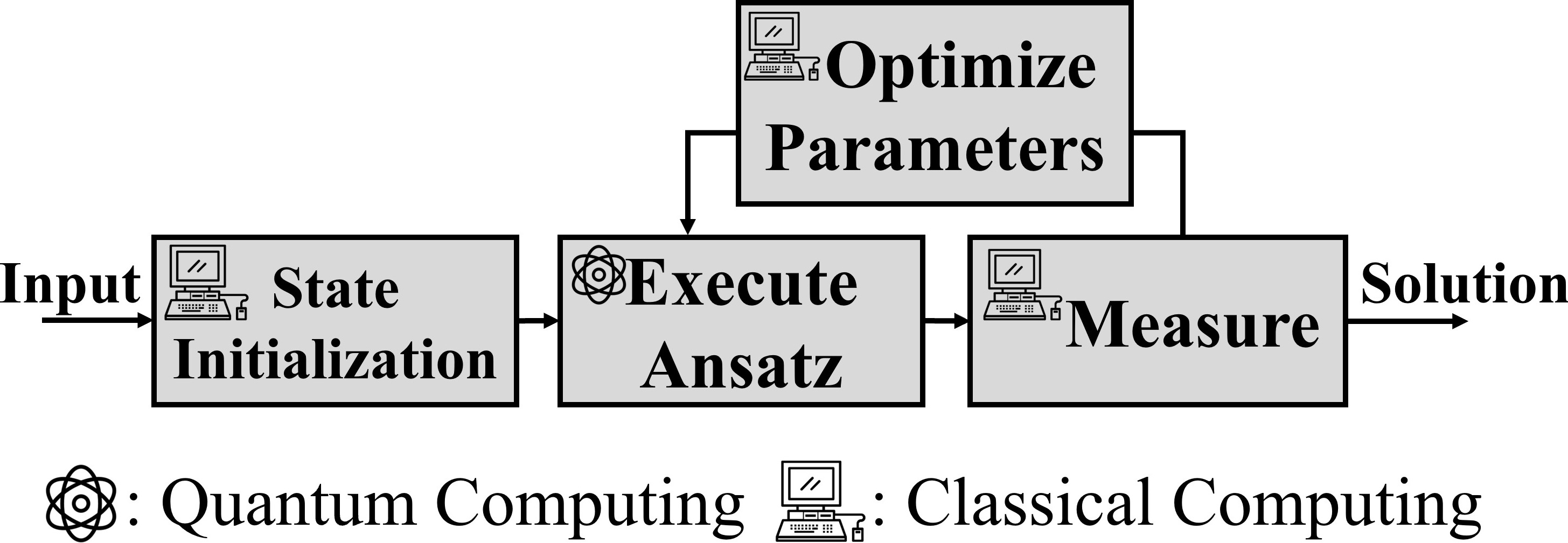}
	\caption{General workflow of a QAOA.}
	\label{fig:qaoa}
\end{figure}

The most representative examples of VQA approaches are the Quantum Approximate Optimization Algorithm (QAOA, \cite{farhi2014quantum}) and the Variational Quantum Eigensolver (VQE, \cite{peruzzo2014variational}). This paper is focused on the former method, the QAOA. We refer readers interested in VQE and QAOA to survey papers such as \cite{fedorov2022vqe} and \cite{blekos2023review}. For illustrative purposes, we depict the general workflow of a QAOA in Figure \ref{fig:qaoa}, based on the work published in \cite{osaba2024hybrid}. 

More specifically, a QAOA is a classical-quantum hybrid algorithm which objective is to leverage the advantages of both computing paradigms. Following Figure \ref{fig:qaoa}, the state initialization for the quantum program is done on a classical computer. This ansatz with parameterized quantum gates is then executed on the quantum computer before its output is transferred back to a classical computer. These parameters are optimized and updated in the quantum program, such that the ansatz, now with updated parameters, can be run again. This process is iterated until the parameters converge.

It is important to note that the development of quantum algorithms may require a considerable level of knowledge related to quantum physics. This complexity may pose a barrier for researchers, especially those coming from computer science, who do not have enough knowledge of fields such as physics or quantum mechanics. With the aim of breaking down that wall and facilitating access to QC for a wider public, several frameworks and programming languages are being proposed, such as the Munich Quantum Toolkit\footnote{\url{https://www.cda.cit.tum.de/research/quantum/mqt/}}
or Silq\footnote{\url{https://silq.ethz.ch/}}. Within this group, we can also find \texttt{Eclipse} \texttt{Qrisp}\footnote{\url{https://qrisp.eu/index.html}}.

The research presented in this paper is focused on \texttt{Eclipse} \texttt{Qrisp}, aiming to demonstrate the efficiency of its QAOA Module. To achieve that, we have conducted an experimentation solving 15 {\it Maximum Cut Problem} (MCP) datasets. We compare the results obtained by \texttt{Eclipse} \texttt{Qrisp}'s QAOA Module with the ones obtained by two different QAOA algorithms implemented using IBM's Qiskit toolkit\footnote{\url{https://www.ibm.com/quantum/qiskit}}.

The remainder of this study is divided into three sections: in Section \ref{sec:qrisp}, \texttt{Eclipse} \texttt{Qrisp} high-level programming language is introduced, specifically focusing on the QAOA module used in the paper. The experimentation carried out is shown in Section \ref{sec:exp}. Conclusions and further work are pointed out in Section \ref{sec:conc}.

\section{\texttt{Eclipse} \texttt{Qrisp} programming language} \label{sec:qrisp}

\texttt{Eclipse} \texttt{Qrisp} is an open-source Python framework for high-level programming of quantum computers. Its unique feature is that it moves away from building quantum algorithms by applying quantum gates directly to the qubits. Instead, it approaches them using variables and functions and thus automates many programming tasks. In \cite{seidel2024qrisp}, the \texttt{Qrisp} framework, with the \texttt{QuantumVariable} and the different quantum types at the core, is described in detail. Furthermore, \cite{seidel2024backtracking} gives an example of how a sophisticated quantum algorithm, in this case the quantum backtracking algorithm, can be implemented in \texttt{Qrisp} and how the framework helps to design quantum algorithms in a manner akin to classical programming.

Another standout feature of \texttt{Eclipse} \texttt{Qrisp} is its modularity, organizing the code into independent modules with minimal interactivity and allowing a team to efficiently manage different sections of the codebase. Such design also facilitates the replacement of modules when enhancements or improved alternatives are proposed. In the context of QAOA, such examples include initialization of QAOA using Trotterized Quantum Annealing (TQA) \cite{sack2021quantum} or recursive QAOA (RQAOA) \cite{bravyi2020obstacles}, among others.

The QAOA module in \texttt{Eclipse} \texttt{Qrisp} is inspired by the Quantum Alternating Operator Ansatz \cite{hadfield2019quantum}, further expanding upon QAOA by introducing a broader range of operators beyond the ones derived in the original paper. In \cite{hadfield2019quantum}, authors formulate various problem instances by defining the cost function, the initial state, the phase separator, and the mixer. They also provide detailed problem formulations, most of which have been implemented in \texttt{Eclipse} \texttt{Qrisp} through the \texttt{QAOAProblem} class. Compared to other frameworks, the use of custom \texttt{QuantumVariables} in \texttt{Eclipse} \texttt{Qrisp} allows flexibility in choosing encoding methods, making it easy to implement otherwise complex problem instances, i.e., the Max-$\kappa$-Colorable Subgraph problem.

Lastly, the source code of \texttt{Eclipse} \texttt{Qrisp} is openly accessible\footnote{https://github.com/eclipse-qrisp/Qrisp}. Furthermore, tutorials on how to solve various optimization problems using the QAOA Module are also available\footnote{https://qrisp.eu/reference/Examples/QAOA.html}.

\section{Experimentation} \label{sec:exp}

\begin{table}[t]
	\centering{
		\caption{Parameterization used for the three QAOAs.}
		\label{tab:params}
		\resizebox{0.45\columnwidth}{!}{
			\begin{tabular}{ll}
				\toprule
				\textbf{Parameter} & Value\\
				\midrule
				\textbf{\# of runs per instance} & 5\\
				\textbf{Max. \# of iterations} & 5.000 \\
				\textbf{\# of shots} & 10.000 \\
				\textbf{\# of layers} & \{1,3,5\} \\
				\textbf{Optimizer} & COBYLA \\
				\bottomrule
			\end{tabular}
	}}
\end{table}

\begin{table*}[t]
	\caption{Results obtained by the QAOAs. For each combination of instance and number of layers, the average and standard deviation of the approximation ratio is represented. In \texttt{bold} the best result per instance and QAOA variant (\textit{1-}, \textit{3-} and \textit{5-layer}).}
	\label{tab:exp}
	\resizebox{1.0\columnwidth}{!}{
		\begin{tabular}{|c|ll|ll|ll|ll|ll|ll|ll|ll|ll|}
			\toprule
			& \multicolumn{6}{|c|}{\bf \texttt{Qiskit-Library QAOA}} & \multicolumn{6}{c|}{\bf \texttt{ad-hoc QAOA}} & \multicolumn{6}{|c|}{\bf Qrisp QAOA Module} \\\midrule
			\multirow{2}{*}{} & \multicolumn{2}{c|}{\textit{1-Layer}} & \multicolumn{2}{c|}{\textit{3-Layer}} & \multicolumn{2}{c|}{\textit{5-Layer}} & \multicolumn{2}{c|}{\textit{1-Layer}} & \multicolumn{2}{c|}{\textit{3-Layer}} & \multicolumn{2}{c|}{\textit{5-Layer}} & \multicolumn{2}{c|}{\textit{1-Layer}} & \multicolumn{2}{c|}{\textit{3-Layer}} & \multicolumn{2}{c|}{\textit{5-Layer}} \\
			& Av. & St. & Av. & St. & Av. & St. & Av. & St. & Av. & St. & Av. & St. & Av. & St. & Av. & St. & Av. & St. \\
			\midrule
			\texttt{MC\_8} & 0.74 & 0.23 & 0.74 & 0.17 & 0.82 & 0.12 & \textbf{0.86} & 0.08 & \textbf{0.88} & 0.10 & \textbf{0.80} & 0.10 & 0.79 & 0.02 & 0.83 & 0.02 & 0.78 & 0.07 \\
			\texttt{MC\_10} & \textbf{0.80} & 0.05 & 0.75 & 0.04 & 0.70 & 0.18 & 0.78 & 0.04 & \textbf{0.81} & 0.07 & \textbf{0.84} & 0.04 & 0.78 & 0.01 & \textbf{0.81} & 0.06 & 0.77 & 0.02 \\
			\texttt{MC\_12} & \textbf{0.85} & 0.18 & 0.77 & 0.19 & 0.79 & 0.09 & 0.82 & 0.02 & 0.81 & 0.08 & 0.78 & 0.18 & 0.84 & 0.01 & \textbf{0.84} & 0.06 & \textbf{0.83} & 0.05 \\
			\texttt{MC\_14} & \textbf{0.81} & 0.09 & 0.82 & 0.08 & \textbf{0.81} & 0.05 & 0.78 & 0.05 & \textbf{0.86} & 0.07 & \textbf{0.81} & 0.04 & \textbf{0.81} & 0.03 & 0.81 & 0.05 & 0.78 & 0.04 \\
			\texttt{MC\_15} & \textbf{0.81} & 0.09 & 0.78 & 0.08 & \textbf{0.80} & 0.05 & 0.80 & 0.02 & \textbf{0.84} & 0.01 & 0.74 & 0.21 & \textbf{0.81} & 0.03 & 0.81 & 0.06 & \textbf{0.80} & 0.04 \\
			\texttt{MC\_16} & \textbf{0.85} & 0.02 & 0.80 & 0.08 & 0.78 & 0.06 & 0.78 & 0.05 & 0.71 & 0.24 & \textbf{0.83} & 0.03 & 0.83 & 0.02 & \textbf{0.81} & 0.03 & 0.82 & 0.03 \\
			\texttt{MC\_17} & \textbf{0.85} & 0.05 & 0.67 & 0.20 & 0.80 & 0.06 & 0.85 & 0.03 & \textbf{0.86} & 0.02 & 0.80 & 0.06 & 0.80 & 0.01 & 0.83 & 0.05 & \textbf{0.81} & 0.04 \\
			\texttt{MC\_18} & 0.78 & 0.06 & 0.76 & 0.12 & 0.76 & 0.11 & 0.74 & 0.02 & 0.76 & 0.04 & 0.75 & 0.05 & \textbf{0.79} & 0.03 & \textbf{0.81} & 0.04 & \textbf{0.80} & 0.04 \\
			\texttt{MC\_19} & 0.80 & 0.09 & 0.80 & 0.10 & 0.77 & 0.11 & 0.82 & 0.03 & 0.83 & 0.02 & 0.84 & 0.05 & \textbf{0.85} & 0.02 & \textbf{0.86} & 0.02 & \textbf{0.85} & 0.04 \\
			\texttt{MC\_20} & 0.79 & 0.12 & 0.82 & 0.06 & 0.68 & 0.08 & 0.77 & 0.08 & 0.82 & 0.06 & 0.78 & 0.03 & \textbf{0.81} & 0.02 & \textbf{0.83} & 0.02 & \textbf{0.80} & 0.03 \\
			\texttt{MC\_21} & \textbf{0.81} & 0.08 & 0.76 & 0.03 & 0.74 & 0.05 & 0.76 & 0.10 & \textbf{0.80} & 0.05 & \textbf{0.80} & 0.04 & 0.78 & 0.03 & \textbf{0.80} & 0.06 & 0.79 & 0.06 \\
			\texttt{MC\_22} & 0.75 & 0.07 & 0.80 & 0.09 & 0.77 & 0.06 & \textbf{0.86} & 0.04 & 0.82 & 0.06 & 0.79 & 0.03 & 0.83 & 0.03 & \textbf{0.85} & 0.05 & \textbf{0.82} & 0.04 \\
			\texttt{MC\_23} & 0.78 & 0.16 & \textbf{0.86} & 0.04 & 0.80 & 0.09 & 0.84 & 0.03 & 0.83 & 0.03 & \textbf{0.84} & 0.06 & \textbf{0.85} & 0.03 & \textbf{0.86} & 0.03 & 0.82 & 0.01 \\
			\texttt{MC\_24} & \textbf{0.85} & 0.05 & 0.80 & 0.01 & 0.64 & 0.16 & 0.82 & 0.03 & \textbf{0.82} & 0.02 & 0.79 & 0.03 & 0.83 & 0.01 & 0.81 & 0.03 & \textbf{0.82} & 0.03 \\
			\texttt{MC\_25} & 0.78 & 0.04 & 0.68 & 0.09 & 0.80 & 0.04 & 0.81 & 0.03 & 0.83 & 0.03 & 0.76 & 0.02 & \textbf{0.83} & 0.02 & \textbf{0.85} & 0.03 & \textbf{0.82} & 0.03 \\    
			\bottomrule
		\end{tabular}
	}
\end{table*}

In the experimentation designed in this paper, \texttt{Eclipse} \texttt{Qrisp}'s QAOA Module is compared with two alternative QAOAs implemented using Qiskit. The first one, coined \texttt{Qiskit- Library QAOA}, resorts to the QAOA library implemented by Qiskit\footnote{\url{https://docs.quantum.ibm.com/api/qiskit/0.28/qiskit.algorithms.QAOA}}. The second algorithm, named \texttt{ad-hoc QAOA}, has been implemented by generalizing a code proposed by IBM\footnote{\url{https://qiskit-rigetti.readthedocs.io/en/v0.4.1/examples/qaoa\_qiskit.html}}. In fact, this code has been used for building the circuits of both Qiskit algorithms.

Table \ref{tab:params} summarizes the parameterization utilized for all three considered methods. Furthermore, regarding the simulators used, \textit{QasmSimulator} has been embraced for Qiskit-based QAOAs, while \textit{Integrated Qrisp Simulator} has been utilized for the \texttt{Eclipse} \texttt{Qrisp}'s QAOA Module.

The performance of the proposed QAOAs has been gauged over 15 MCP datasets, randomly generated using the Python script introduced in \cite{osaba2023qoptlib}. The size of the considered datasets is between 8 and 25 nodes. For building the corresponding QUBOs, the \textit{MaxCut} open library included in Qiskit v0.6.0 has been employed\footnote{\url{https://qiskit.org/ecosystem/optimization/stubs/qiskit\_optimization.applications.Maxcut.html}}. As depicted in Table \ref{tab:params}, five independent executions have been run for each (problem, technique) combination, aiming to provide statistically reliable findings on the performance of every technique.

The results obtained are represented in Table \ref{tab:exp}. We depict in that table the average and standard deviation of the approximation ratio (AR) obtained by each method. More specifically, the AR has been calculated using as a reference the optimum values of each instance, which have been obtained using the industry-oriented Quantagonia's Hybrid Solver\footnote{\url{https://www.quantagonia.com/hybridsolver}}. For the sake of replicability, the instances, results, and algorithms implemented are openly available\footnote{\url{https://doi.org/10.17632/b5gbz44m99.1}}.

\begin{figure}[t]
	\centering
	\includegraphics[width=0.7\linewidth]{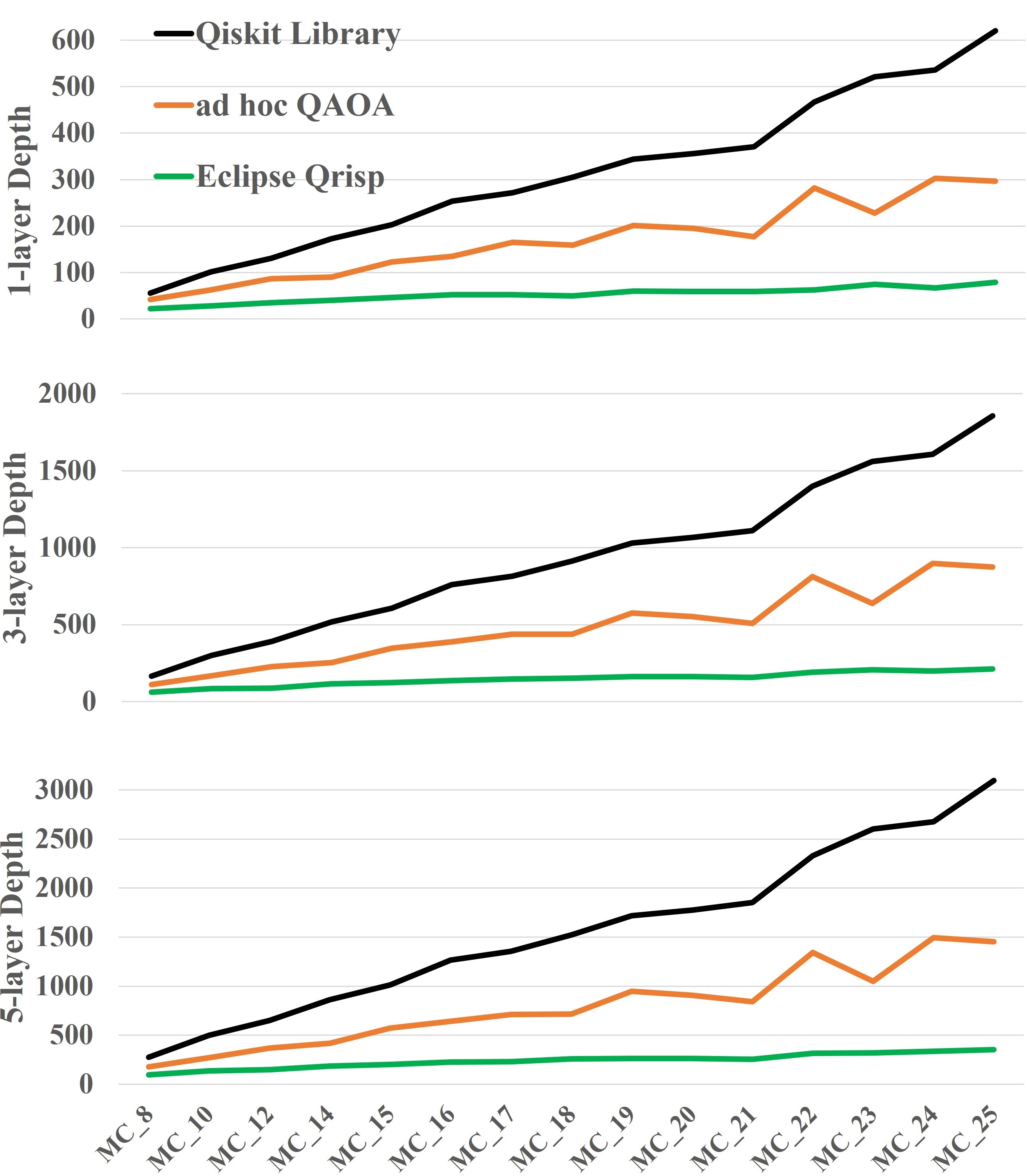}
	\caption{Compiled Quantum Circuit Depths of \texttt{Qiskit-Library QAOA}, \texttt{ad-hoc QAOA} and \texttt{Eclipse} \texttt{Qrisp}'s QAOA.}
	\label{fig:depths}
\end{figure}

\section{Discussion and future work} \label{sec:conc}

In summary, the results obtained from this work clearly demonstrate the promising performance of the \texttt{Eclipse} \texttt{Qrisp}'s QAOA Module. In fact, it has emerged as the best alternative for both the \textit{3-layer} and \textit{5-layer} variants, securing the best results in 26 out of 45 comparisons.

The importance of these results is accentuated by the fact that the depths of the circuits employed by \texttt{Eclipse} \texttt{Qrisp}'s QAOA are significantly better than \texttt{Qiskit-Library QAOA} and \texttt{ad-hoc QAOA}. We represent these depths in Figure \ref{fig:depths} for all the QAOA variants implemented in this research. As seen from the figure, the increase associated with \texttt{Eclipse} \texttt{Qrisp} is markedly less rapid in contrast to other methods, indicating its superior efficiency.

In short, the existence of languages such as \texttt{Eclipse} \texttt{Qrisp} contributes to building a multidisciplinary community around quantum computing \cite{villar2023hybrid}. This will undoubtedly help the field progress towards new, as yet unknown, horizons. As part of future work, more thorough experimentation has been planned. This includes performing rigorous statistical tests, benchmarking against a broader set of problem instances, and tackling new instances of the Maximum Cut Problem.

\section*{Acknowledgments}
This research was funded by the European Union, project OASEES (HORIZON-CL4-2022, grant agreement no 101092702), and the by the Federal Ministry for Economic Affairs and Climate Action (German: Bundesministerium für Wirtschaft und Klimaschutz), project Qompiler (grant agreement no: 01MQ22005A).

\bibliographystyle{splncs04}
\bibliography{biblio.bib}
\end{document}